\documentclass{ws-rv975x65}
\usepackage{subfigure}     
\usepackage[square]{ws-rv-van}     
\makeindex

\newcommand{\GeV}{\,\mbox{GeV}}
\newcommand{\gev}{\rm{GeV}}

\def\Vub  {\it{|V_{ub}|}}
\def\Vcb  {\it {|V_{cb}|}}
\def\as  {\alpha_s}
\def \be{\begin{equation}}
\def \ee{\end{equation}}
\newcommand{\bea}{\begin{eqnarray}}
\newcommand{\eea}{\end{eqnarray}}
\def \nn{\nonumber}
\newcommand\gsim{\mathop{\mbox{\vbox{\hbox{$>$} \vskip -9pt \hbox{$\sim$}
             \vskip -3pt  }}}}

\begin{document}

\chapter[Inclusive semileptonic $B$ secays]{Inclusive semileptonic $B$ decays and $|V_{cb}|$\\In memoriam Kolya Uraltsev }\label{ra_ch1}

\author[Paolo Gambino]{Paolo Gambino 
}

\address{Universit\`a di Torino, Dipartimento di Fisica, and INFN, Torino\\
Via Giuria 1, I-10125 Torino, Italy \\
gambino@to.infn.it
}

\begin{abstract}
The magnitude of the CKM matrix elements $V_{cb}$  can be extracted from inclusive 
semileptonic $B$ decays in a model independent way pioneered by Kolya Uraltsev and collaborators. I review here the present status and latest developments in this field.

\end{abstract}

\body

\newcommand{\mupi}{\mu_\pi^2}
\newcommand{\mug}{\mu_G^2}
\newcommand{\rd}{\rho_D^3}
\newcommand{\rls}{\rho_{LS}^3}
\section{A semileptonic collaboration}
My continuing involvement with semileptonic $B$ decays  is mostly due to a fortuitous encounter with Kolya Uraltsev in 2002. A group of experimentalists of the DELPHI Collaboration, among whom Marco Battaglia and Achille Stocchi, had embarked in an analysis of semileptonic moments and asked Kolya and me to help them out. He was the expert, I was a novice in the field and things have stayed that way for a long time thereafter.
The joint paper that appeared later that year \cite{Battaglia:2002tm} contained one of the first 
fits to semileptonic data to extract $|V_{cb}|$, the masses of the heavy quarks and some non-perturbative parameters, and it  used  Kolya's proposal to  avoid any $1/m_c$ expansion \cite{Uraltsev:2002ta}.
 The next step for us was to 
compute the moments with a cut on the lepton energy \cite{Gambino:2004qm}, as measured by Cleo and Babar \cite{Mahmood:2004kq,Aubert:2004tea}. Impressed as I was by Kolya's deep physical insight and enthusiasm, I was glad that he asked me to continue our collaboration. 
Kolya thought that a global fit should be performed by  experimentalists, but as 
theoretical  issues kept arising  he tirelessly discussed with them every single detail;
the  BaBar fit \cite{Aubert:2004aw}, where   the kinetic scheme analysis of \cite{Battaglia:2002tm} was extended to the BaBar dataset, and the global fit of \cite{Buchmuller:2005zv}
owe very much to his determination.

Kolya's patience in explaining  
was unlimited and admirable: countless times I took advantage of it and learned from him. Our semileptonic collaboration later covered perturbative corrections \cite{Aquila:2005hq}, the extraction of $|V_{ub}|$ \cite{Gambino:2007rp}, and a reassessment of the zero-recoil heavy quark sum rule \cite{Gambino:2010bp,Gambino:2012rd}.  It was during one of his visits to Turin that he suffered a first heart attack, but it did not take long before he was back to his usual dynamism.
Working with Kolya was sometimes complicated, but it was invariably rewarding. He was stubborn and 
we could passionately argue about a single point for hours. 
For him discussion, even heated discussion,  was an essential part of doing physics.
I will forever miss Kolya's passionate love of physics and his total dedication  to science. They were the marks of a noble soul, a kind and discreet friend. 

In the following I will review the present status of the inclusive $B\to X_c \ell\bar\nu$ decays, the subject of most of my work with Kolya, who was a pioneer of the field. Semileptonic $B$ decays allow for a precise determination of the magnitude of the 
CKM matrix elements $V_{cb}$ and $V_{ub}$, which are in turn crucial ingredients in the 
analysis of CP violation in the quark sector and in the prediction of  flavour-
changing neutral current transitions. In the case of inclusive decays, the Operator Product Expansion (OPE) allows us to 
describe the relevant non-perturbative physics in terms of a finite number of non-perturbative parameters that can be extracted from experiment, while in the case of exclusive decays 
like $B\to D^{(*)}\ell \bar \nu$ or $B\to \pi\ell \bar \nu$ the form factors have to be computed by non-perturbative methods, {\it e.g.}\ on the lattice.
Presently,  the most precise determinations of $|V_{cb}|$
(the inclusive one \cite{Alberti:2014yda} and the one based on $B\to D^* \ell\nu$ at zero recoil and a lattice calculation of the form-factor \cite{Bailey:2014tva}) show a $\sim3\sigma$ discrepancy that does not seem to admit a new physics explanation, as I will explain later on.
A similar discrepancy between the inclusive and exclusive determinations occurs in the 
case of $|V_{ub}|$ \cite{Amhis:2012bh}. It is a pity that Kolya  will not witness
how things eventually settle.

\section{The framework}
Our understanding of inclusive semileptonic $B$ decays is based on a
simple idea: since inclusive decays sum over all possible 
hadronic final states, the quark in 
the final state hadronizes with unit probability and the 
transition amplitude is sensitive only to the long-distance dynamics  
of the initial $B$ meson. Thanks to the large hierarchy between the typical 
energy release, of $O(m_b)$, and the hadronic scale  
$\Lambda_{\rm QCD}$, and to asymptotic freedom, any residual
sensitivity to non-pertur\-bative effects is suppressed by powers of
$\Lambda_{\rm QCD}/m_b$.

The OPE allows us to express the
nonperturbative physics in terms of $B$ meson matrix elements of
local operators of dimension $d\ge 5$, while the Wilson coefficients
can be expressed as a perturbative series in $\alpha_s$
\citep{Chay:1990da,Bigi:1992su,Bigi:1993fe,Blok:1993va,Manohar:1993qn}. The OPE
disentangles the physics associated with {\it soft} scales of order
$\Lambda_{\rm QCD}$ (parameterized by the matrix elements of the local
operators) from that associated with {\it hard} scales $\sim m_b$, which determine 
the Wilson coefficients. The total semileptonic width and the moments
of the kinematic distributions are therefore double expansions 
in  $\alpha_s$ and  $\Lambda_{\rm QCD}/m_b$,
with a leading term that is given by the free $b$ quark decay. 
Quite importantly, the power corrections start at $O(\Lambda^2_{\rm
  QCD}/m_b^2)$ and are comparatively suppressed.
At higher orders in the OPE,   terms suppressed by 
powers of $m_c$ also appear, starting with $O(\Lambda_{\rm QCD}^3 /
m_b^3 \times \Lambda_{\rm QCD}^2 / m_c^2)$ \citep{Bigi:2009ym}.
For instance, the expansion for the total semileptonic width is
\bea
\Gamma_{sl} &=&\! \Gamma_0 \Big[ 1+a^{(1)} \frac{\as(m_b)}{\pi}\! +a^{(2,\beta_0)}\beta_0\!\left(\frac{\as}{\pi}\right)^2 \!\!+a^{(2)}\!\!\left(\frac{\as}{\pi}\right)^2 \!  \nn
\\ &&\!+
\left(\!-\frac12  +p^{(1)}  \, \frac{\as}{\pi}\right)\nonumber
 \frac{\mu_\pi^2}{m_b^2}   + \left(g^{(0)} +g^{(1)}\, \frac{\as}{\pi}\right)\frac{\mu_G^2(m_b)}{m_b^2}  \\
&& \left. \ \ \ \ \ \ \
+
 d^{(0)} \frac{\rho_D^3}{m_b^3} -g^{(0)} \frac{\rho_{LS}^3}{m_b^3} + {\rm higher\  orders}
  \right], \label{width}
\eea
where $\Gamma_0= A_{ew}\,|V_{cb}^2| \,G_F^2 \,m_b^5 \,(1-8\rho+8\rho^3 -\rho^4 -12\rho^2 \ln\rho)/192 \pi^3$ is the tree level free quark decay width, $\rho=m_c^2/m_b^2$, and $A_{ew}\simeq 1.014$  the leading electroweak correction. I have split the 
  $\as^2$ coefficient into a BLM piece proportional to  $\beta_0=11-2/3 n_f$  and a remainder. 
The expansions for the moments have the same structure. 

The relevant parameters in the double series of Eq.\,(\ref{width}) are the heavy quark
masses $m_b$ and  $m_c$, the strong coupling $\alpha_s$, and the
$B$ meson expectation values of local operators of dimension 5 and 6, denoted by $\mupi, \mug, \rd,\rls$. 
As there are only two
dimension five operators, two matrix elements appear at $O(1/m_b^2)$:
\begin{eqnarray}
  \mupi (\mu)  &=& \frac1{2M_B}\langle B |  \bar b_v\,\vec \pi^2 \,b_v|B
  \rangle_\mu, \\
  \mug(\mu) &=& \frac1{2M_B}\langle B| \bar b_v\frac{i}2 \sigma_{\mu\nu}
  G^{\mu\nu} b_v|B\rangle_\mu
\end{eqnarray} 
where $\vec \pi= -i \vec D$, $D^\mu$ is the covariant derivative, 
$b_v(x)=e^{-i m_b v\cdot x}b(x)$ is the $b$ field deprived of its high-frequency modes, and
$G^{\mu\nu}$ the gluon field tensor. The matrix element of the kinetic
operator, $\mu_\pi^2$, is naturally associated
with the average kinetic energy of the $b$ quark in the $B$ meson, while
that of the chromomagnetic operator, $\mug$, is related to the
$B^*$-$B$ hyperfine mass splitting. They generally
depend on  a cutoff $\mu=O(1 \gev)$  chosen to separate soft and
hard physics. The cutoff can be implemented in different ways. In the
kinetic scheme \citep{Bigi:1996si,Bigi:1994ga}, a Wilson cutoff on
the gluon momentum is employed in the $b$ quark rest frame: all soft
gluon contributions are attributed to the expectation values of the
higher dimensional operators, while hard gluons with momentum $|\vec
k|>\mu$ contribute to the perturbative corrections to the Wilson
coefficients. 
Most current applications of the OPE involve 
$O(1/m_b^3)$ effects \citep{Gremm:1996df} as well, parameterized in
terms of two additional parameters, generally denoted by  $\rd$ and
$\rls$ \citep{Bigi:1994ga}. 
All of the OPE parameters describe universal properties of the $B$ meson or of the
quarks and are useful in several applications.

The interesting quantities to be measured are the total rate and some
global shape parameters, such as the mean and variance  of the lepton
energy spectrum or of the hadronic invariant mass distribution.  As most experiments can detect the leptons only above a certain threshold in energy, the
lepton energy moments are defined as
\be
  \langle
  E^n_\ell\rangle=\frac{1}{\Gamma_{E_\ell>E_\mathrm{cut}}}\int_{E_\ell>E_\mathrm{cut}}
   E^n_\ell\ \frac{d\Gamma}{dE_\ell} \ dE_\ell~,\label{eq3}
\ee
where $E_\ell$ is the lepton energy in $B\to X_c\ell\nu$,
$\Gamma_{E_\ell>E_\mathrm{cut}}$ is the semileptonic width above the energy
threshold $E_\mathrm{cut}$ and $d\Gamma/dE_\ell$ is the differential
semileptonic width as a function of $E_\ell$. The hadronic mass
moments are
\be
  \langle
  m^{2n}_X\rangle=\frac{1}{\Gamma_{E_\ell>E_\mathrm{cut}}}\int_{E_\ell>E_\mathrm{cut}}
  m^{2n}_X\ \frac{d\Gamma}{dm^2_X}\ dm^2_X~.\label{eq4}
\ee
Here, $d\Gamma/dm^2_X$ is the differential width as a function of the squared
mass of the hadronic system $X$. For both types of moments, $n$ is the
order of the moment. For $n>1$, the moments can also be defined
relative to $\langle E_\ell\rangle$ and $\langle m^2_X\rangle$,
respectively, in which case they are called central moments:
\be
\ell_1(E_{cut})=\langle E_\ell \rangle_{E_\ell > E_{cut}}, \quad\quad \quad
\ell_{2,3}(E_{cut})=\langle \left(E_\ell - \langle E_\ell \rangle \right)^{2,3} \rangle_{E_\ell > E_{cut}}\,;
\ee
\be
h_1(E_{cut})=\langle M_X^2\rangle_{E_\ell > E_{cut}}, \quad \quad
h_{2,3}(E_{cut})=\langle (M_X^2-\langle M_X^2\rangle)^{2,3}\rangle_{E_\ell > E_{cut}} .
\ee
Since the physical information that can be extracted from the first three linear moments 
is highly correlated, it is more convenient to study the central moments $\ell_i$ and $h_i$, which correspond to the mean, variance, and asymmetry of the lepton energy and invariant mass distributions.

The OPE cannot be expected to converge in regions of phase space where 
the momentum of the final hadronic state is $O(\Lambda_{\rm QCD})$ and
where perturbation theory has singularities. This is because what
actually controls the expansion is not $m_b$ but the energy release, which is
$O(\Lambda_{\rm QCD})$ in those cases.
The OPE is therefore valid only for sufficiently inclusive
measurements and in general cannot  describe differential
distributions.
The lepton energy moments can be measured very
precisely, while the hadronic mass central moments are directly sensitive to
higher dimensional matrix elements such as $\mu_\pi^2$ and $\rho_D^3$. 
The leptonic and hadronic moments, which are independent of $\Vcb$, give us
constraints on the quark masses and on the non-perturbative OPE matrix
elements, which can then be used, together with additional information, in the total semileptonic width  to extract  $\Vcb$.

\section{Higher order effects}
The reliability of the inclusive method depends on our ability to
control the higher order contributions in the double series and to
constrain quark-hadron duality violation, i.e.\ effects beyond the OPE,
which we know to exist but expect to be rather suppressed in semileptonic
decays. The calculation of higher order effects allows us to verify
the convergence of the double series and to reduce and properly
estimate the residual theoretical uncertainty. Duality violation, see \citep{Bigi:2001ys} for a review, is related to the analytic continuation of the OPE to Minkowski space-time. It
can be constrained {\em
  a posteriori}, considering how well the OPE predictions fit the
experimental data. This in turn  depends on precise
measurements and precise OPE predictions. As the experimental accuracy
reached at the $B$ factories is already better than the theoretical accuracy for
most of the measured moments and will further improve at Belle-II,   efforts to improve the latter are strongly motivated.

The main ingredients for an accurate analysis of the experimental data
on the moments and the subsequent extraction of $\Vcb$ have been
known for some time. Let us consider first  the purely
perturbative contributions. The $O(\as)$ perturbative
corrections to various kinematic distributions and to the rate have
been computed long ago. In particular,  the complete $O(\as)$  and $O(\as^2 \beta_0)$ corrections to the charged 
leptonic spectrum have been first calculated in  \cite{Jezabek:1988iv,Jezabek:1988ja,Czarnecki:1994pu} and \cite{Gremm:1996gg}. 
The so-called BLM corrections \cite{Brodsky:1982gc},  of $O(\alpha_s^2
\beta_0)$, are related to the running of the strong coupling inside the loops and are usually the dominant source of two-loop
corrections in $B$ decays.
The first $O(\as)$ calculations of the hadronic spectra
appeared in  \citep{Czarnecki:1989bz,Falk:1995me,Falk:1997jq} and were later completed 
in \citep{Trott:2004xc,Uraltsev:2004in,Aquila:2005hq}, 
while the  $O(\as^2\beta_0)$  contributions were studied in
\cite{Falk:1995me,Falk:1997jq,Uraltsev:2004in,Aquila:2005hq}.
The triple differential distribution was
first computed at $O(\as)$ in \citep{Trott:2004xc,Aquila:2005hq}; its   $O(\as^n \beta_0^{n-1})$  corrections  can be found  in \citep{Aquila:2005hq}. 

The complete two-loop perturbative corrections
to the width and moments of the lepton energy and hadronic mass
distributions have been  computed in 
\citep{Pak:2008qt,Melnikov:2008qs,Biswas:2009rb} by both numerical and
analytic methods. The kinetic scheme implementation for actual
observables  can be found in \citep{Gambino:2011cq}.
In general, using $\as(m_b)$ in the one-loop result and adopting the on-shell scheme for the quark masses, the non-BLM
corrections amount to about $-20\%$ of the two-loop BLM  corrections
and give small contributions to normalized moments. In the kinetic
scheme with cutoff $\mu=1$\gev, the perturbative expansion of the total
width is
\begin{eqnarray}
  \Gamma[\bar{B} \to X_c e \bar{\nu}] &\propto& 1 - 0.96\,
  \frac{\as(m_b)}{\pi} -0.48\, \beta_0 \left( \frac{\as}{\pi}
  \right)^2 \nonumber\\ &&
  + 0.82 \left( \frac{\as}{\pi} \right)^2 + O(\as^3) \approx
  0.916 \label{expkin}
\end{eqnarray}
Higher order BLM corrections of $O(\as^n \beta_0^{n-1})$ to the width
are also known \citep{Benson:2003kp,Aquila:2005hq} and can be resummed in the kinetic scheme: the
resummed BLM result is numerically very close to that of from NNLO calculations
\citep{Benson:2003kp}. The residual perturbative error in the total
width is  about 1\%.

In the normalized leptonic moments the perturbative
corrections cancel to a large extent, independently of the mass scheme, 
because hard gluon emission is comparatively suppressed. This pattern of
cancelations, crucial for a correct estimate of the theoretical
uncertainties, is confirmed by the complete $O(\alpha_s^2)$ calculation,
although the numerical precision of the available results is not
sufficient to improve the overall accuracy for the higher central
leptonic moments \citep{Gambino:2011cq}. The non-BLM corrections turn
out to be  more important for the hadronic moments. Even though it improves
the overall theoretical uncertainty only moderately, the complete NNLO
calculation leads to the meaningful inclusion of precise mass
constraints in
various perturbative schemes.

The coefficients of the non-perturbative corrections of  $O(\Lambda^n_{\rm QCD}/m_b^n)$ in the double series are Wilson coefficients of 
power-suppressed local operators and can be computed perturbatively.
The calculation of the $O(\alpha_s\Lambda_{\rm QCD}^2/m_b^2)$ corrections has been recently completed. 
The $O(\as)$ corrections to  the coefficient of  $\mupi$ have been computed numerically in 
\cite{Becher:2007tk} and analytically in \cite{Alberti:2012dn}. They   can be also obtained 
from the  parton level $O(\as)$ result using reparameterization invariance (RI) relations 
\cite{Bigi:1992su,Luke:1992cs,Manohar:2010sf}.  In fact, these RI relations have represented a useful check for 
the  calculation of the remaining 
$O(\alpha_s\Lambda_{\rm QCD}^2/m_b^2)$  corrections, those proportional to $\mu^2_G$,
which was completed in \cite{Alberti:2013kxa}.
\begin{figure}[t]
\begin{center}
\includegraphics[width=10.5cm]{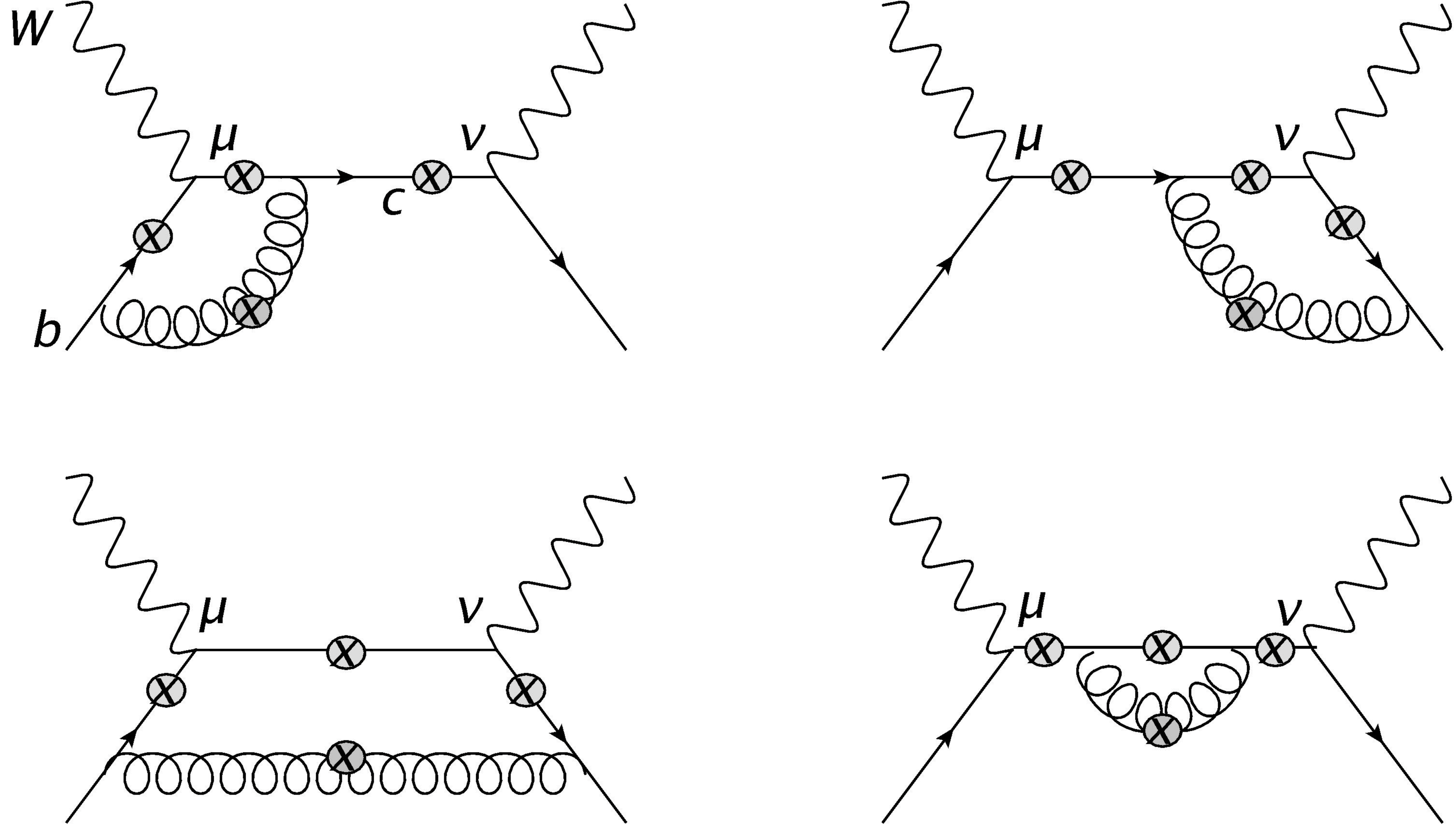}
\caption{\sf One-loop diagrams contributing to the current correlator. The background gluon can be attached wherever a  cross is marked.}
\label{fig1}
\end{center}
\end{figure}
The calculation consists in matching the one-loop diagrams in Fig.~\ref{fig1}, representing the correlator of two axial-vector currents computed in an expansion around the mass-shell of the $b$ quark,  onto local HQET operators. 

 A recent  independent calculation \cite{Mannel:2014xza} of the semileptonic width at $m_c=0$  seems to be in agreement with the $m_c\to 0$ limit of \cite{Alberti:2013kxa}.
Refs. \cite{Alberti:2012dn,Alberti:2013kxa} provide analytic results for the $O(\alpha_s\Lambda_{\rm QCD}^2/m_b^2)$ corrections to the three relevant structure functions and hence to  the triple differential 
semileptonic $B$ decay width. The most general moment
have now been  computed to this order and  employed 
to improve the precision of the  fits to $|V_{cb}|$ \citep{Alberti:2014yda}.

Numerically, using for  the heavy quark on-shell masses the  values
$m_b=4.6$ GeV and $m_c=1.15$ GeV,  the total semileptonic width reads
\be
\Gamma_{B\to X_c \ell \nu}=\Gamma_0\left[ \left(1-1.78\, \frac{\as}{\pi}\right)\left(1-\frac{\mu_\pi^2}{2m_b^2}\right) - \left(1.94 +2.42  \,\frac{\as}{\pi}\right)\frac{\mu_G^2(m_b)}{m_b^2} \right],\nonumber
\ee
where $\Gamma_0$
is the tree level width  and we have omitted higher order terms of $O(\as^2)$ and $O(1/m_b^3)$. The coefficient of $\mu_\pi^2$ is fixed by RI (or equivalently, by Lorentz invariance) at all orders. 
The parameter $\mug$ is renormalized at the scale $m_b$.  It is advisable to evaluate the QCD coupling constant at a scale lower than $m_b$. If we adopt  \(\alpha_s = 0.25 \) the $O(\alpha_s )$ correction increases the \(\mu_G^2 \) coefficient by about  7\%. In the kinetic scheme with 
 cutoff $\mu=1$GeV and for the same values of the masses the width becomes
\be
\Gamma_{B\to X_c \ell \nu}=\Gamma_0 \left[ 1-0.96\, \frac{\as}{\pi} -  \left(\frac12- 0.99\, \frac{\as}{\pi}\right)\frac{\mu_\pi^2}{m_b^2} - \left(1.94 +3.46  \,\frac{\as}{\pi}\right)\frac{\mu_G^2(m_b)}{m_b^2} \right],\nn 
\ee
where the NLO corrections to the coefficients of $\mupi,\mug$ are both close to 15\% but have different signs.
  Overall, the $O(\as\Lambda^2_{\rm QCD}/m_b^2)$ contributions 
decrease the total width by about 0.3\%. However, NLO corrections also modify the 
coefficients of  $\mupi,\mug$ in the moments which are fitted to extract the non-perturbative parameters, and will ultimately shift the values of the OPE parameters to be employed in the width. Therefore, in order to quantify the eventual numerical impact of the new corrections on the semileptonic width and on $|V_{cb}|$, a new global fit has to be performed.
\begin{figure}[t]
\begin{center}
\includegraphics[width=7cm]{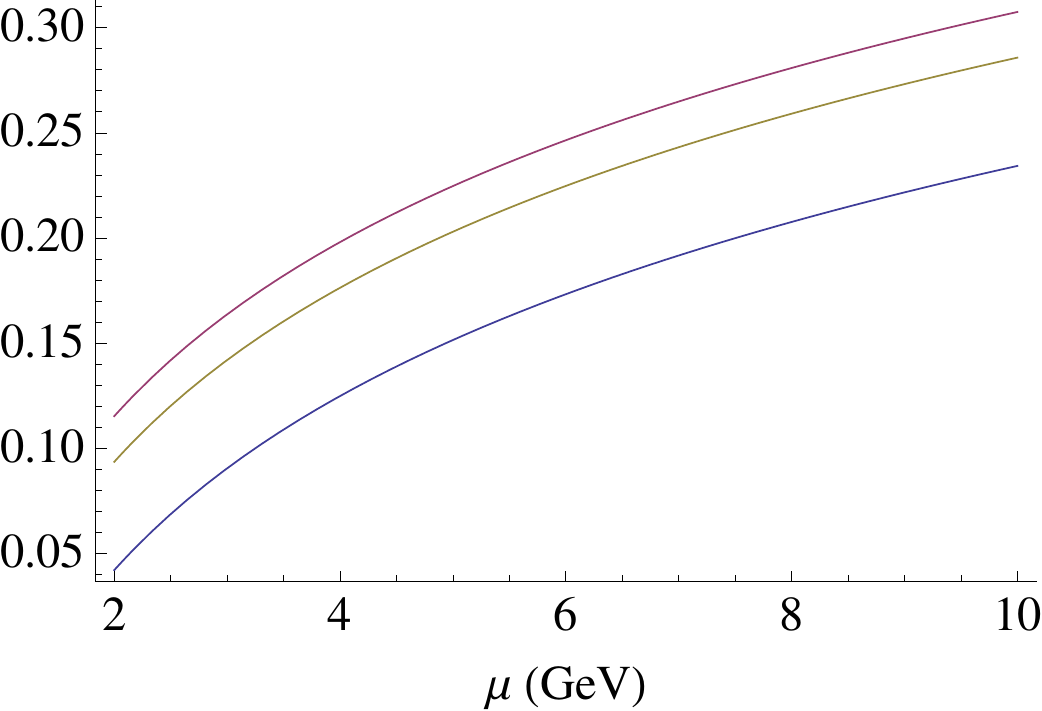}
\caption{\sf Relative NLO correction to the $\mug$ coefficients in the width (blue), first (red) and second central (yellow) leptonic moments as a function of the renormalization scale  of $\mug$.}
\label{fig2}
\end{center}
\end{figure}
The size of the $O(\as \mug/m_b^2)$ corrections depends on the renormalization scale $\mu$ of the chromomagnetic operator. This is illustrated in Fig.\,\ref{fig2}, where the size of the NLO correction relative to the tree level results is shown for the width and the first two leptonic central moments at different values of $\mu$. The NLO corrections are quite small for $\mu\approx 2$GeV and, as expected,  increase with $\mu$. For $\mu\gsim m_b$ the running of $\mug$ appears to  dominate  the NLO corrections.
In view of the importance of $O(1/m_b^3)$ corrections, if a
theoretical precision of 1\% in the decay rate is to be reached, the
$O(\as/m_b^3)$ effects need to be calculated.

As to the higher  power corrections, the
 $O(1/m_b^4)$ and $O(1/m_Q^5)$ effects were computed in
\citep{Mannel:2010wj}. The main problem here is the proliferation of
non-perturbative  parameters:  as many as nine new
expectation values appear at $O(1/m_b^4)$ and more at the next
order. Because they cannot all be extracted from experiment, in
\citep{Mannel:2010wj}
they have been estimated in the ground state 
saturation approximation, thus
reducing them to products of the known $O(1/m_b^{2,3})$
parameters, see also \citep{Heinonen:2014dxa}.
 In this approximation, the total $O(1/m_Q^{4,5})$
correction to the width  is about +1.3\%. The $O(1/m_Q^5)$ effects are
dominated by $O(1/m_b^{3} m_c^2)$ intrinsic charm contributions,
amounting to  +0.7\% \citep{Bigi:2009ym}. The net effect on
$\Vcb$ also depends  on the corrections to the moments. Ref.~\cite{Mannel:2010wj} estimate that the overall effect on $\Vcb$ is
a 0.4\% increase. While this sets the scale of  higher order
power corrections, it is as yet unclear how much the result depends on
the assumptions made for the expectation values. A new preliminary 
global fit \cite{Turzcyk}  performed using different ansatz for the  new non-perturbative parameters
seems to confirm that these corrections lead to a small shift in $\Vcb$. 

Two implementations of the OPE calculation have been  employed in
global analyses; they are based either on the kinetic scheme
\citep{Bigi:1996si,Bigi:1994ga,Benson:2003kp,Gambino:2004qm} or on the $1S$ mass scheme for the $b$ quark mass
\citep{Hoang:1998hm,Bauer:2004ve}. They both include power corrections up to and including 
$O(1/m_b^3)$ and perturbative corrections of  $O(\alpha_s^2
\beta_0)$. 
Beside differing in the perturbative scheme adopted, the global fits
may include a different choice of experimental data, employ specific
assumptions, or estimate the theoretical uncertainties in different 
ways. 
Recently, the kinetic scheme  implementation has been upgraded to include first the complete $O(\as^2)$ \cite{Gambino:2011cq} and later the $(\as \Lambda^2/m_b^2)$ \cite{Alberti:2014yda} contributions.


\section{$|V_{cb}|$ and the fit to semileptonic moments}

The OPE parameters can be constrained by 
various moments of the lepton energy and hadron mass distributions of $B\to X_c \ell \nu$ 
that have been   measured with good accuracy at the $B$-factories, as well as at CLEO, DELPHI, CDF \cite{Csorna:2004kp,Aubert:2004aw,Aubert:2009qda,Urquijo:2006wd,Schwanda:2006nf,Acosta:2005qh,Abdallah:2005cx}.
The total semileptonic width can then be employed to extract  $|V_{cb}|$. The situation is 
less favorable in the case of $|V_{ub}|$, where the total rate is much more difficult to access experimentally because of the background from $B\to X_c \ell \nu$,
but the results of the semileptonic fits are  crucial also in that case.
This strategy has been rather successful and has allowed for a $\sim 2\%$ determination of
$V_{cb}$ and for a  $\sim 5\%$ determination of
$V_{ub}$ from inclusive decays \cite{Amhis:2012bh,Bevan:2014iga}. 

The first few moments of the charged lepton energy spectrum in   $B\to X_c\ell \nu$ decays are experimentally measured with high precision ---
better than 0.2\% in the case of the first moment. 
At the $B$-factories a lower cut on the lepton energy, $E_\ell \ge E_{cut}$, is  applied to suppress the  background. Experiments measure  the moments at different values of $E_{cut}$, which provides additional information as the cut dependence is also a function of the OPE parameters. The relevant quantities are therefore $\ell_{1,2,3}$, $h_{1,2,3}$,
as well as the ratio $R^*$ between the rate with and without a cut
\begin{equation}
R^* (E_{cut})= \label{eq:Rstar}
\frac{\int_{ E_{cut}}^{E_{max}} d E_\ell \ \frac{d\Gamma}{d E_\ell}}
{\int_0^{E_{max}} d E_\ell \ \frac{d\Gamma}{d E_\ell}}\ .
\end{equation}
This quantity is needed to relate the actual measurement of the rate with a cut to the total rate, from which one conventionally extracts $|V_{cb}|$. 
All of these observables can be expressed as  double expansions in $\alpha_s$ and 
inverse powers of $m_b$, schematically
\bea
M_i&=& M_i^{(0)}+ \frac{\alpha_s(\mu)}{\pi} M_i^{(1)}+ \left(\frac{\alpha_s}{\pi}\right)^2 M_i^{(2)} + \left(M_i^{(\pi,0)}+  \frac{\alpha_s(\mu)}{\pi}M_i^{(\pi,1)}\right)
 \frac{\mu_\pi^2}{m_b^2} \nonumber\\
&&+\left(M_i^{(G,0)}+ \frac{\alpha_s(\mu)}{\pi}M_i^{(G,1)}\right)
 \frac{\mu_G^2}{m_b^2}+ M_i^{(D)} \frac{\rho_D^3}{m_b^3}+ M_i^{(LS)} \frac{\rho_{LS}^3}{m_b^3} +\dots
\label{double}
\eea
where all the coefficients $M_i^{(j)}$ depend on $m_c$, $m_b$,  
  $E_{cut}$, and on various renormalization scales. 
The dots represent 
missing terms of $O(\alpha_s^3)$, $O(\alpha_s^2/m_b^2)$, $O(\alpha_s/m_b^3)$, and $O(1/m_b^4)$, which are either 
 unknown or not yet included in the latest analysis \cite{Alberti:2014yda}.
It is worth stressing that according to the adopted definition  the OPE parameters $\mu_\pi^2$, ... are matrix elements of local  operators 
evaluated in the physical $B$ meson, {\it i.e.}\ without taking the infinite mass limit.

\begin{figure}[t]
\begin{center}
\includegraphics[width=6.2cm]{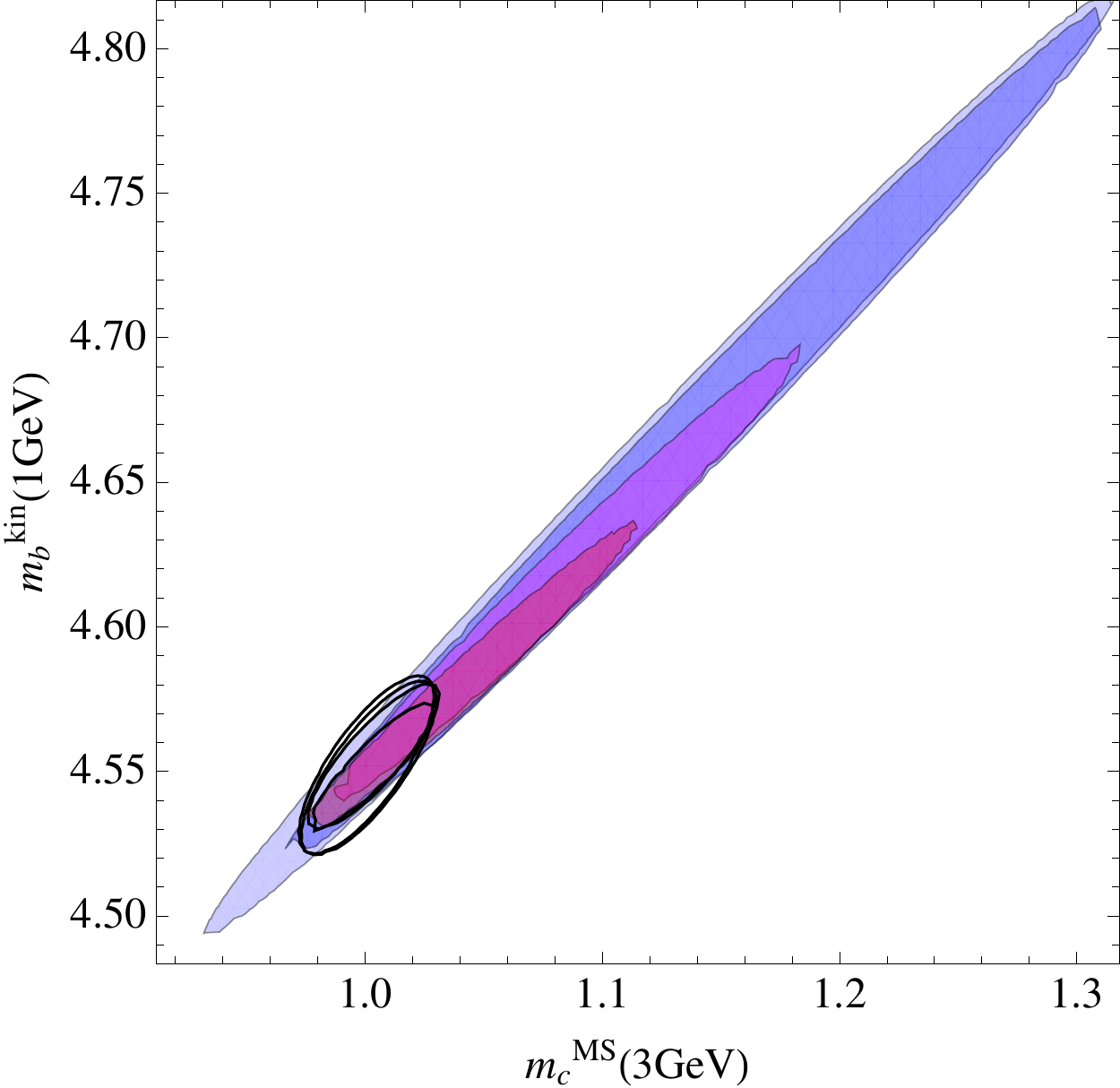}
\includegraphics[width=6.2cm]{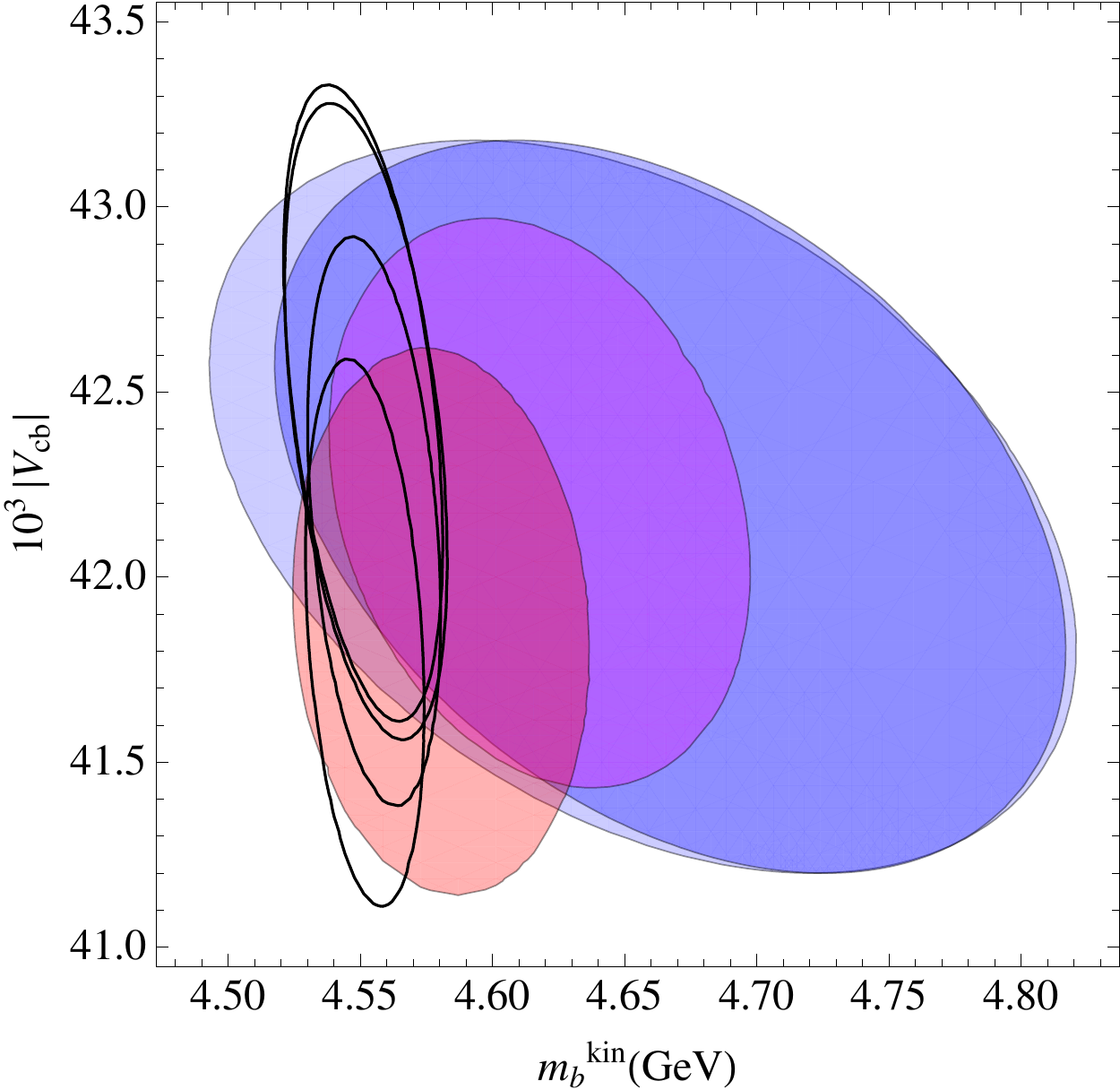}
\caption{\sf Two-dimensional projections of the fits performed with different assumptions for the theoretical correlations. The orange, magenta, blue, light blue 1-sigma regions correspond to the four scenarios considered in \cite{Gambino:2013rza}. The black contours show  the same regions when the $m_c$ constraint of Ref.~\cite{Dehnadi:2011gc} is employed.
}
\label{fig3}
\end{center}
\end{figure}

The semileptonic moments are sensitive to a specific
linear  combination of  $m_c$ and $m_b$, $\approx m_b-0.8 m_c$ \citep{Voloshin:1994cy}, see Fig.~\ref{fig3},
which is close to the one needed for the extraction of
$\Vcb$,  but they cannot resolve the individual 
masses with good accuracy.
 It is important to check
the consistency of the constraints on $m_c$ and $m_b$ from 
semileptonic moments with  precise determinations 
of these quark masses, as a step in the effort to improve
our theoretical description of inclusive semileptonic decays. Moreover, the
inclusion of these constraints in the semileptonic fits  
improves the accuracy of the $\Vub$ and $\Vcb$
determinations. 
The heavy quark
masses and the non-perturbative parameters obtained from the fits are also relevant for a precise
calculation of other inclusive decay rates such as that of $B\to
X_s\gamma$ \citep{Gambino:2013rza}.

In the past,  the first two moments of the photon energy  in  $B\to X_s \gamma$ have generally been employed to improve the accuracy of the fit. Indeed,   the first moment corresponds to a  determination of $m_b$.
However, in recent years  
rather precise determinations of the heavy quark masses ($e^+e^-$ sum rules, lattice QCD etc.) have 
become  available, based on completely different methods, see {\it e.g.} \cite{Chetyrkin:2009fv,Dehnadi:2011gc,Bodenstein:2011ma,Signer:2008da,Allison:2008xk,McNeile:2010ji,Blossier:2010cr,Heitger:2013oaa,Carrasco:2014cwa,Lucha:2013gta} and  \cite{Antonelli:2009ws,Beringer:1900zz} for reviews.  The charm mass determinations have a smaller absolute uncertainty and appear quite consistent with each other, providing a good external 
constraint for the semileptonic fits.
Radiative moments remain interesting in their own respect, 
but  they are not competitive with the charm mass determinations. Moreover, 
experiments place
a lower cut on the photon energy, which introduces a sensitivity 
to the Fermi motion
of the $b$-quark inside the $B$ meson and tends to disrupt the OPE. One
can still resum the higher-order terms into a non-local distribution
function 
and parameterize it
assuming different functional forms \cite{Neubert:1993um,Bigi:1993ex,Benson:2004sg}, but the parameterization will depend on $m_b, \mupi$ etc., namely the same parameters one wants to extract. Another
serious problem is that only the leading operator contributing to
inclusive radiative decays can be described by an OPE.  
Therefore, radiative moments are in 
principle subject to additional $O(\Lambda_{QCD}/m_b)$ effects, which have not yet been 
estimated \cite{Paz:2010wu}.  
For all these reasons the most recent analyses \cite{Gambino:2013rza,Alberti:2014yda} have relied solely on 
charm and possibly bottom mass determinations.

 The global fits of Refs.~\cite{Gambino:2013rza,Alberti:2014yda} are performed in the 
kinetic scheme with a cutoff $\mu=1\GeV$ and follow the implementation described in \cite{Gambino:2004qm,Gambino:2011cq}.  The two fits only differ in the inclusion of $O(\as^2/m_b^2)$ corrections 
and in the consequent reduction of theoretical uncertainties.
In order to use  the high precision $m_c$ determinations
without introducing additional 
theoretical uncertainty due to the mass scheme conversion,
it is convenient to employ the $\overline{\rm MS}$ scheme for the 
charm mass, denoted by $\overline{m}_c(\mu_c)$, and to choose  a  normalization scale $\mu_c$ well above $m_c$, {\it e.g.}\ 3 \GeV.

The experimental data for the moments are fitted to the theoretical expressions in order to 
constrain the 
non-perturbative parameters and the heavy quark masses.  43  measurements are included, see \cite{Gambino:2013rza} for the list.
 The chromomagnetic expectation value $\mu_G^2$ is also constrained 
by the hyperfine splitting
\be 
M_{B^*}-M_B= \frac23 \frac{\mu_G^2}{m_b} + O\left(\frac{\as\mug}{m_b},\frac1{m_b^2}\right)\nonumber.
\ee 
Unfortunately, little is known of the power
corrections to the above relation and only a loose bound \cite{Uraltsev:2001ih} can be set, see \cite{Gambino:2012rd} for a recent discussion. 
For what concerns $\rho_{LS}^3$, it is somewhat constrained by the heavy quark sum rules \cite{Uraltsev:2001ih}. Refs.~\cite{Gambino:2013rza,Alberti:2014yda} use the constraints 
\be
\mu_G^2=(0.35\pm 0.07)\GeV^2,
\qquad \rho_{LS}^3=(-0.15\pm 0.10 )\GeV^3\, . \label{constraints}
\ee
It should be stressed that $\rls$ plays a minor role in the fits because its coefficients are 
generally suppressed with respect to the other parameters.

It is interesting to note that the fit without theoretical uncertainties is not  good, 
 with   $\chi^2/dof\sim 2$, corresponding to a very small $p$-value and driven by a strong tension ($\sim 3.5 \sigma$) between the constraints in Eq.~(\ref{constraints}) and the measured moments. On the other hand, the fit without the constraints (\ref{constraints})
 is not too bad.
Indeed, theoretical uncertainties are not so much necessary for the OPE expressions to fit the moments --- that would merely test Eq.(\ref{double}) as a parameterization;   they are instead needed to preserve the definition  of the parameters as $B$ expectation values of 
certain local operators,  which in turn can  be employed in the semileptonic widths and 
in other applications of the Heavy Quark Expansion.

As noted above,  the OPE description of semileptonic moments is subject to two sources of theoretical 
uncertainty: missing higher order terms in 
Eq.~(\ref{double}) and terms that violate  quark-hadron duality.  Only of the first kind of uncertainty is usually considered: 
the violation of local quark-hadron duality would manifest itself as an inconsistency of the fit, which as we will see 
is certainly not present at the current level of theoretical and experimental accuracy.

In \cite{Alberti:2014yda} we assume that missing  perturbative corrections can affect the coefficients of $\mupi$ 
and $\mu_G^2$  at the level of $\pm 7\%$, while missing perturbative 
 and higher power corrections can effectively change the coefficients of $
\rho_D^3$ and $\rho_{LS}^3$ by $\pm 30\%$. Moreover we assign an irreducible theoretical 
uncertainty of 8 MeV to the heavy quark masses, and vary $\alpha_s(m_b)$ by 0.018. 
The changes in $M_i$ due to these variations of the fundamental parameters are added in quadrature and provide a theoretical uncertainty $\delta M_i^{th}$, to be subsequently added 
in quadrature with the experimental one, $\delta M_i^{exp}$.
This method is consistent with the residual scale dependence observed at NNLO, and appears to be reliable: the NNLO corrections and the $O(1/m_b^{4,5})$ (using 
ground state saturation as in \cite{Mannel:2010wj}) have been found to be within the range 
of expectations based on the method in the original formulation of \cite{Gambino:2004qm}. 

The correlation between  theoretical errors assigned to different observables is much 
harder to estimate, but  plays an important role in the semileptonic fits. Let us first consider moments computed at a fixed value of $E_{cut}$: as 
long as one deals with central higher moments, there is no argument of principle supporting a 
correlation between two different moments, for instance $\ell_1$ and $h_2$. We also do not observe 
any clear pattern in the known corrections, and therefore regard the theoretical predictions 
for  different central moments as completely uncorrelated.
Let us now consider the calculation of a certain moment $M_i$ for two close values of $E_{cut}$, say 1\GeV\ and 1.1\GeV. Clearly, the OPE expansion for $M_i(1\GeV)$ will
be very similar to the one for $M_i(1.1\GeV)$, and we may expect this to be true 
at any order in $\as$ and $1/m_b$. The theoretical uncertainties we assign to $M_i(1\GeV)$
and $M_i(1.1\GeV)$ will therefore be very close to each other and {\it very highly 
correlated}. The degree of correlation between the theory uncertainty of $M_i(E_1)$ and 
$M_i(E_2)$ can intuitively be expected to decrease as $|E_1-E_2|$ grows.
Moreover, we know that higher power corrections are going to modify significantly the 
spectrum only close to the endpoint. Indeed, one observes  that the $O(1/m_b^{4,5})$ 
contributions are equal for all cuts below about 1.2\GeV\ (see Fig.2 of \cite{Mannel:
2010wj}) and the same happens for the $O(\as\mupi/m_b^2)$ corrections \cite{Becher:
2007tk}. Therefore, the dominant sources of current theoretical uncertainty
suggest very high correlations among the theoretical predictions of the moments for cuts
below roughly 1.2 \GeV.

Various assumptions on the theoretical correlations have been tried. A 
100\% correlation between a certain central moment computed at different values of $E_{cut}$ has been assumed {\it e.g.}\ in \cite{Buchmuller:2005zv}). This is too strong an assumption, which ends up distorting the fit 
because the dependence of $M_i$ on $E_{cut}$, itself a function of 
the fit parameters, is  then free of theoretical uncertainty. A fit performed in this way will underestimate the uncertainties. Another possibility has been proposed in
 Ref.~\cite{Bauer:2004ve}, with the theoretical correlation matrix equal to the experimental one.

 Four alternative approaches for the theoretical correlations are compared in \cite{Gambino:2013rza}. 
 Fig.~\ref{fig3} shows some of the results of the fits performed with the four options for the 
 theoretical correlations. The fits include  the two 
 constraints of Eq.~(\ref{constraints}). 
 In general, the results depend sensitively on the option adopted. In the case of the heavy 
quark masses, which are strongly correlated, we observe  large errors differing significantly between the various options, although the central values are quite consistent.  
The results for the non-perturbative parameters depend even stronger on the 
option.
The inclusion of  precise mass constraints
in the fit decreases the errors and neutralizes the  ambiguity
due to the ansatz for the theoretical correlations. It also allows us to check the consistency 
of the results with independent information. 
The effect of the inclusion of a precise charm mass constraint in the semileptonic fit is illustrated in 
Fig.~\ref{fig3}. As expected, the 
uncertainty in the $b$ mass becomes about 20-25MeV in all scenarios, a marked improvement, also with respect to the precision resulting from the use of radiative moments \cite{Amhis:2012bh}.
The inclusion of the $m_c$ constraint indeed stabilized the fits with respect to the ansatz for the theory correlations.
On the other 
hand, there is hardly any  improvement  in the final precision of the non-perturbative 
parameters and of $|V_{cb}|$. 

\begin{figure}[t]
\begin{center}
\includegraphics[width=5.4cm]{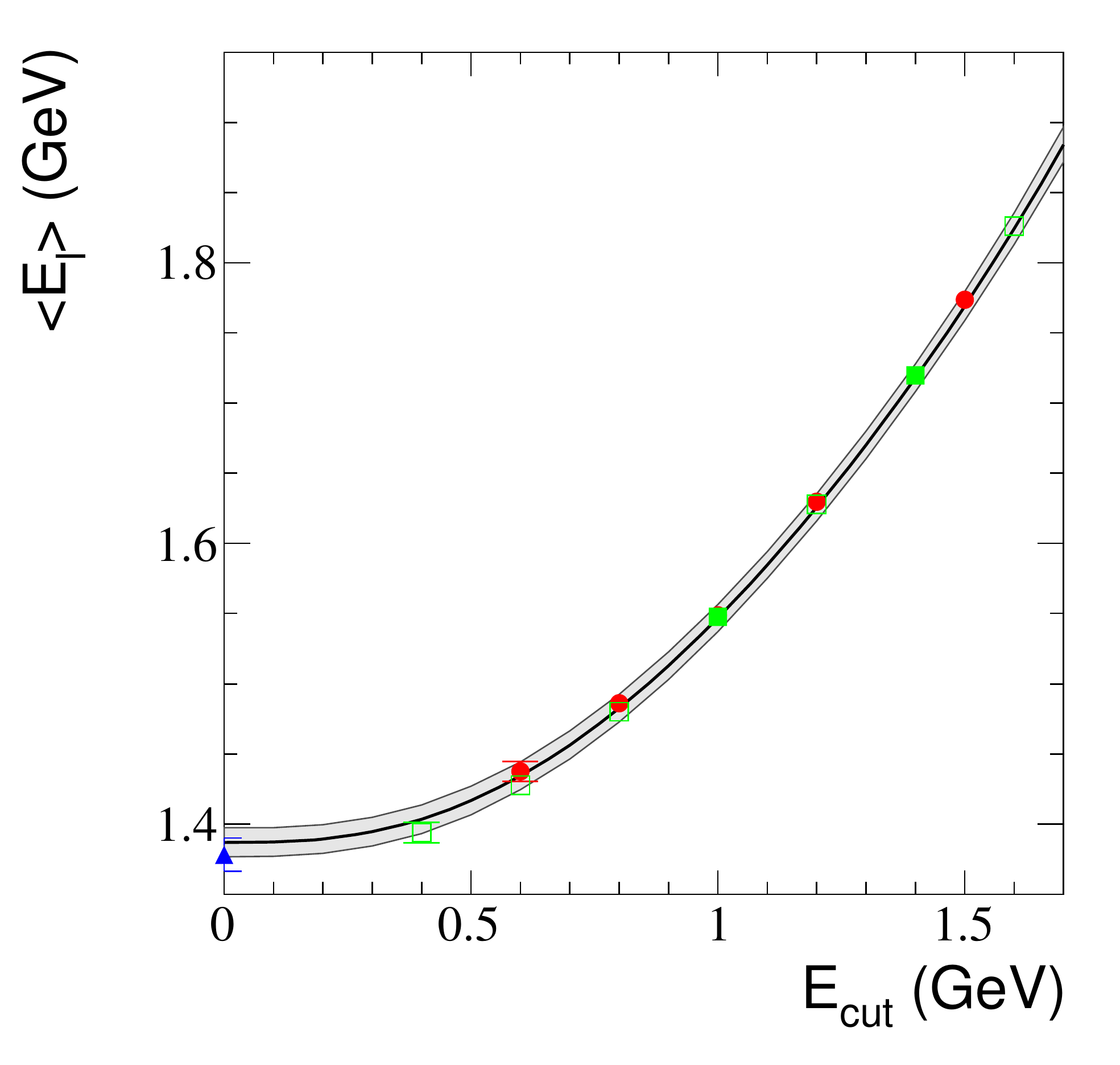}\ \ \
\includegraphics[width=5.4cm]{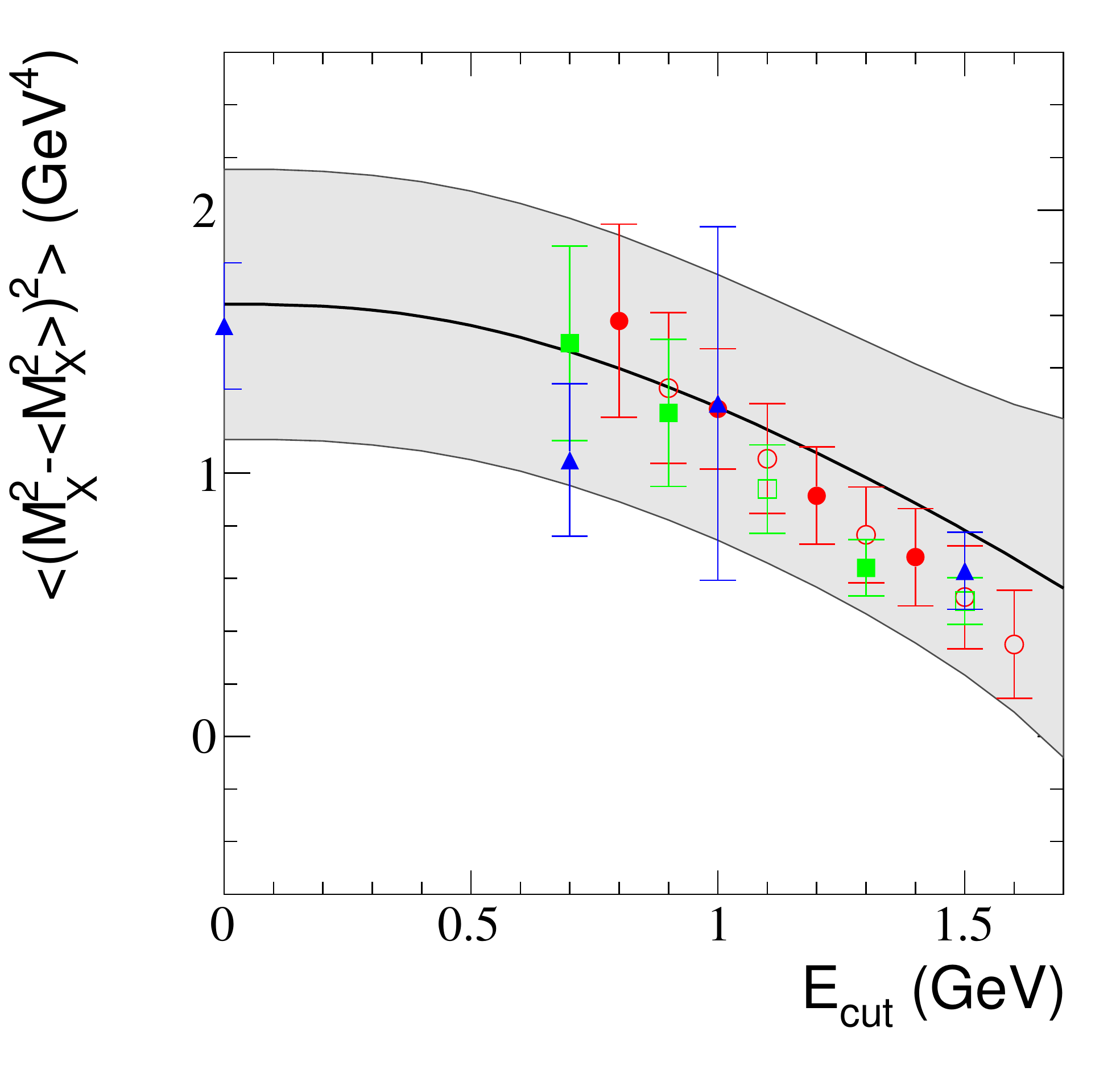}\vspace{-4mm}
\caption{\label{plots}  \sf Default fit predictions for $ \ell_{1}$ and $h_2$
compared with measured values  as  functions of $E_{cut}$.
 The grey band is the theory prediction with total theory error.  Filled symbols mean that the point was used in the fit. Open symbols are measurements that were not used in the fit.
 BaBar data are shown by circles, Belle by squares and other experiments (DELPHI, CDF, CLEO) by triangles.
}
\end{center}
\end{figure}

Figs.~\ref{plots}  show two examples of leptonic and hadronic moments measurements compared with their theoretical prediction based on the results of the fit of \cite{Gambino:2013rza} with theory uncertainty.  As anticipated, theory errors are generally larger than experimental ones. The situation is similar also for the fits in \cite{Alberti:2014yda}.

\begin{figure}[t]
\begin{center}
\includegraphics[width=8.6cm]{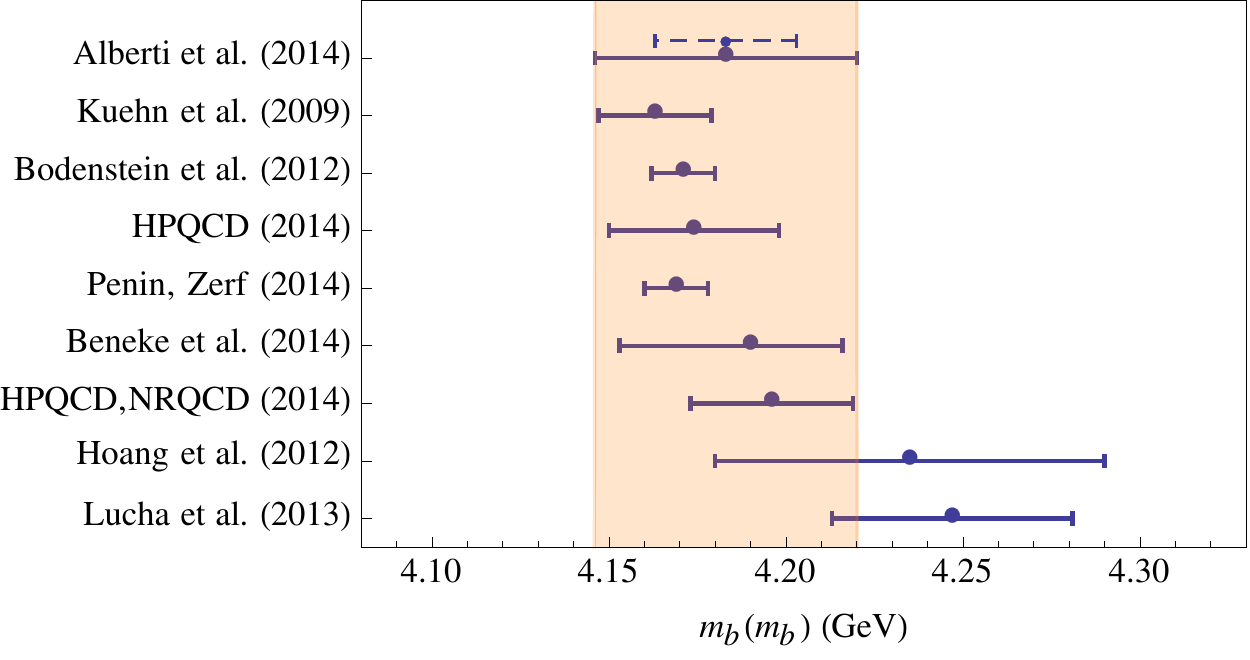}
\caption{ \sf \label{mb} Comparison of different $\overline m_b(\overline m_b)$ determinations \cite{Alberti:2014yda,Chetyrkin:2009fv,Bodenstein:2011fv, Beneke:2014pta,Colquhoun:2014ica,Chakraborty:2014aca,Hoang:2012us, Penin:2014zaa,Lucha:2013gta}. The dashed line denotes the error before scheme conversion.}
\end{center}
\end{figure}

The default fit of \cite{Alberti:2014yda} uses
$\overline m_c(3\GeV)=0.986(13)$ GeV \cite{Chetyrkin:2009fv};
the results are shown in  Table \ref{table:1},
\begin{table}[t]
\tbl{ Results of the global fit  with $m_c$ constraint in the default scenario of \cite{Alberti:2014yda}.
All parameters are in $\GeV$ at the appropriate power and all, except $m_c$, in the kinetic scheme at $\mu=1\GeV$. The last row gives the uncertainties. }
{ \begin{tabular}{@{}cccccccc@{}}\toprule
$m_b^{kin}$ & $ \overline{m_c}(3\GeV)$   &  $\mupi $ &$\rd$ &$\mug$ & $\rls$  & ${\rm BR}_{c\ell\nu}${ (\%)}& $10^3 \,|V_{cb}|$ \\ 
\colrule
    4.553 &0.987 & 0.465 & 0.170 & 0.332 & -0.150 & 10.65 & 42.21 \\  
    0.020 & 0.013 & 0.068 & 0.038 & 0.062 & 0.096 & \ 0.16 & \ 0.78
   \\  \botrule  
\end{tabular} }\label{table:1}
\end{table}
where  the bottom mass is expressed in the kinetic scheme,  $m_b^{kin}(1\GeV)$.
Most available $m_b$ determinations, however, use the $\overline{\rm MS}$  mass
$\overline m_b(\overline m_b)$ which is not  well-suited to the description of semileptonic $B$ decays as the calculation of the width and moments in terms of $\overline m_b(\overline m_b)$ involves large  higher order corrections. 
Since the relation between the kinetic and the $\overline{\rm MS}$  masses is known only to 
$O(\as^2)$, the ensuing uncertainty is not negligible. It has been estimated to be about 30 MeV \cite{Gambino:2011cq},
\be\label{transl}
m_b^{kin}(1\GeV)-\overline m_b(\overline m_b) = 0.37\pm 0.03 \GeV,\nonumber
\ee
leading to a preferred value
\be
\overline m_b(\overline m_b) = 4.183\pm 0.037  \GeV,
\nonumber\ee
in good agreement with various recent $m_b$ determinations
\cite{Chetyrkin:2009fv,Bodenstein:2011fv, Beneke:2014pta,Colquhoun:2014ica,Chakraborty:2014aca,Hoang:2012us, Penin:2014zaa,Lucha:2013gta}, as illustrated in Fig.\ref{mb}.
Of course, one can 
 also include in the fit both $m_c$ and $m_b$ determinations, 
but because of the scheme translation error in $m_b$ the gain in accuracy is  limited \cite{Gambino:2013rza,Alberti:2014yda}.

As already noted,  the semileptonic moments are highly sensitive to a linear 
combination of the heavy quark masses. 
When no external constraint is imposed on $m_{c,b}$, the semileptonic moments determine best  a linear combination of the heavy quark masses which is close to their difference,
\be
m_b^{kin}(1\GeV)-  0.85\, \overline m_c(3\GeV)= 3.714\pm 0.018\GeV, \label{massdiff}.
\ee


The value of $|V_{cb}|$ is computed using
\be
|V_{cb}|=\sqrt{\frac{|V_{cb}|^2\, {\rm BR}_{c\ell \nu}}{\tau_B \,\Gamma_{B\to X_c \ell \nu}^{OPE}} ,
}\ee
with $\tau_B=1.579(5)$ps \cite{Amhis:2012bh}.
Its theoretical error is computed combining in quadrature the parametric uncertainty that results from the fit, and an additional 1.3\% theoretical error
to take into account missing higher order corrections in the expression for the semileptonic 
width. It turns out that using $\overline{m}_c(2\GeV)$ rather than $\overline{m}_c(2\GeV)$  
leads to a better converging expansion for the width, with smaller theoretical error for $|V_{cb}|$, about 1\%. However, the value of $|V_{cb}|$ extracted in this way, $42.01(68)\times 10^{-3}$ is compatible with that in Table 1. The same holds if the kinetic scheme is used also for 
$m_c$, in which case  we obtain $|V_{cb}|=42.04(67)\times 10^{-3}$.


The fits are generally  good, with  $\chi^2/d.o.f. \approx 0.4 $  for the default fit.
The low $\chi^2$ of the default fit is due  to the large theoretical uncertainties we 
have assumed.
It may be tempting to interpret it as evidence that the theoretical errors 
have been overestimated. However,  higher order corrections may 
effectively shift the parameters of the $O(1/m_b^2)$ and  $O(1/m_b^3)$ contributions. If we 
want to maintain the formal definition of these parameters, and to be able to use them 
elsewhere, we therefore have to take into account the potential shift they may experience 
because of higher order effects. 

The fits with a constraint on $m_c$ are quite stable with respect to a  change of inputs. In particular, small differences are found when experimental data at high $E_{cut}$ are excluded,
and when  only hadronic or leptonic moments are considered. 
One may also wonder whether the inclusion of moments measured at
different values of $E_{cut}$ really benefits the final accuracy. It turns out that  the benefit is minor but non-negligible. 

In the kinetic scheme the inequalities
$\mu_\pi^2(\mu)\!\ge\!\mu_G^2(\mu)$, $\rd(\mu)\!\ge\!-\rls(\mu)$
hold at arbitrary values of the cutoff $\mu$. The central values of the fit
satisfy the inequalities. The dependence of the results on the scales of $\as$ and on the kinetic cutoff has been studied in \cite{Alberti:2014yda}. Changing the scale of $\as$ from $m_b$ to 2\GeV\ the value of $|V_{cb}|$ increases by only 0.5\%, well within its error, while $m_b$ increases by less than 0.4\%.



\section{Conclusions}
We have seen that the most recent value of $|V_{cb}|$ extracted from an analysis of inclusive 
semileptonic $B$ decays is \cite{Alberti:2014yda}
\be
|V_{cb}|= (42.21\pm 0.78)\times 10^{-3}\label{incl}.
\ee
 A competitive  
determination of $|V_{cb}|$ derives from a comparison of the extrapolation of the $B\to D^* l \nu$ rate to the zero-recoil point with an unquenched lattice QCD calculation of the zero recoil form factor by the Fermilab-MILC collaboration \cite{Bailey:2014tva},
\be
|V_{cb}|= (39.04\pm 0.49_{exp}\pm 0.53_{lat}\pm 0.19_{ QED})\times 10^{-3}\label{excl},
\ee
where the error is split into experimental, lattice and QED components.
The values in Eqs.\,(\ref{incl},\ref{excl})  disagree by $2.9\sigma$. This is  a long-standing tension, which has become stronger with recent improvements in the OPE and lattice calculations. A few comments are in order:
\begin{itemize}
\item
The zero-recoil form factor of  $B\to D^* l \nu$ can also be estimated using heavy quark sum rules, see \cite{Gambino:2010bp,Gambino:2012rd} for a recent reanalysis.
Although this method is subject to larger uncertainties and it is difficult to improve its accuracy, it leads to a $|V_{cb}|$ compatible with the inclusive determination, $|V_{cb}|=40.93(1.11)\times 10^{-3}$.
There exist also less precise determinations of $|V_{cb}|$ based on  the decay $B\to D l \nu$, but they  do not help resolving the issue at the moment, see \cite{Bevan:2014iga} for a review.

\item The extrapolation of the $B\to D^* l \nu$ experimental data to the zero-recoil point is performed by the experimental collaborations using the Caprini-Lellouch-Neubert parameterization \cite{Caprini:1997mu}, based on HQET at next-to-leading order and expected to reproduce the form factor within 2\% (not included in the present error budget).  While this rigid  parameterization with only two free parameters fits well the experimental data at $w\neq 1$, at the present level of precision its use to extrapolate the rate to zero recoil is questionable. Lattice calculations of the form factors at  non-zero recoil are currently under way; they would allow us to avoid the extrapolation.

\item
It is also possible to determine $|V_{cb}|$ indirectly, using the CKM unitarity relations together with CP violation and flavor data, without the above direct information: SM analyses of this kind   
by the UTfit and CKMFitter collaborations give $(42.05\pm 0.65)\times 10^{-3}$ \cite{Bona:2006ah} and $(41.4^{+2.4}_{-1.4})\times 10^{-3}$  \cite{Charles:2004jd}, 
    both closer to the inclusive value of Eq.\,(\ref{incl}).

\end{itemize}

In principle, the  discrepancy between the values of $|V_{cb}|$ extracted from inclusive decays and from $B\to D^* l \nu$ could be ascribed to physics beyond the SM, 
as the $B\to D^*$ transition is sensitive only to the axial-vector component of the $V-A$ charged weak current. However, the new physics effect should be sizable (8\%),
and 
it would require new interactions which seem  ruled out by electroweak  constraints on the effective $Zb\bar b$ vertex \cite{Crivellin:2014zpa}.
The most likely explanation of the discrepancy between Eqs.\,(\ref{incl},\ref{excl}) is therefore a problem in the theoretical and/or experimental analyses of semileptonic decays.

I do not have space here to discuss the closely related determination of $|V_{ub}|$ from inclusive semileptonic $B$ decays without charm, which shows a similar puzzling tension between inclusive and exclusive determinations. However, I cannot avoid stressing the central 
role played by Kolya from the beginning also in this field  \cite{Bigi:1993ex,Bigi:1997dn,Uraltsev:1999rr,Gambino:2007rp}. The interested reader is referred to \cite{Bevan:2014iga} for a recent review.
\\
\\

This work is supported in part by MIUR under contract 2010YJ2NYW 006,  by the EU Commission through the HiggsTools Initial Training Network PITN-GA-2012-316704,   and by Compagnia di San Paolo under contract ORTO11TPXK.

\bibliographystyle{ws-rv-van}

\end{document}